\documentclass[aps,prb,showpacs,twocolumn]{revtex4}
\usepackage{graphicx}
\begin{document}
\title{Radiation-induced magnetotransport in high-mobility two-dimensional systems:\\
Role of electron heating}
\author{X.L. Lei and S.Y. Liu}
\affiliation{Department of Physics, Shanghai Jiaotong University,
1954 Huashan Road, Shanghai 200030, China}

\begin{abstract}
Effects of microwave radiation on magnetoresistance are analyzed in a 
balance-equation scheme that covers regimes of inter- and intra-Landau level processes 
and takes account of photon-asissted electron transitions as well as radiation-induced 
change of the electron distribution for high mobility two-dimensional systems. 
Short-range scatterings due to background impurities and defects
are shown to be the dominant direct contributors to the photoresistance oscillations.
The electron temperature characterizing the system heating due to irradiation,
is derived by balancing the energy absorption from the radiation field and 
the energy dissipation to the lattice through realistic electron-phonon couplings, 
exhibiting resonant oscillation. Microwave modulations of Shubnikov de Haas 
oscillation amplitude are produced together with microwave-induced resistance 
oscillations, in agreement with experimental findings. In addition, the suppression of 
the magnetoresistance caused by low-frequency radiation in the higher magnetic field 
side is also demonstrated. 

\end{abstract}


\maketitle

\section{Introduction}
The discovery of microwave induced magnetorersistance oscillations (MIMOs) 
and zero-resistance states (ZRS) in high-mobility two-dimensional (2D) 
electron gas (EG)\cite{Zud01,Ye,Mani,Zud03} has stimulated  
tremendous experimental\cite{Yang,Dor,Mani04,Will,Zud04,Stud,Kovalev,Mani-apl,Du04,Dor04} 
and theoretical\cite{Ryz,Ryz86,Anderson,Koul,Andreev,Durst,Xie,Dmitriev,Lei03,Lei04,
Ryz03,Vav,Mikh,Dietel,Torres,Dmitriev04,Inar,Ryz0411,Joas} interest in  
radiation related magneto-transport in 2D electron systems. Since theoretically it
has been shown that the ZRS can be the result of the 
instability induced by absolute negative resistivity,\cite{Andreev} 
the majority of microscopic models focus mainly on MIMOs in spatially uniform cases 
and identify the region where an negative dissipative magnetoresistance
develops as that of measured zero resistance.  
Most of previous investigations concentrated on the range of low magnetic fields 
$\omega_c/\omega \leq 1$ ($\omega_c$ stands for the cyclotron frequency) 
subject to a radiation of frequency $\omega/2\pi\leq 100$\,GHz, 
where MIMOs show up strongly and Shubnikov-de Haas oscillations (SdHOs) 
are rarely appreciable. 
In spite of the fact that both MIMOs and SdHOs are 
magnetoresistance related phenomena appearing in overlapping 
field regimes, little attention was paid to the influence 
of a microwave radiation on SdHO until a recent experimental finding
at higher frequency.\cite{Kovalev}
Further observations clearly show that the amplitudes of SdHOs 
are strongly affected by microwave radiations of different frequency  
in both low ($\omega_c/\omega < 1$) and high ($\omega_c/\omega > 1$) 
magnetic field ranges.\cite{Mani-apl,Du04,Dor04}
Kovalev {\it et al.}\cite{Kovalev} observed a suppression of 
the SdHOs around cyclotron resonance $\omega_c\sim \omega$ 
induced by a radiation of 285\,GHz.
Du {\it et al.}\cite{Du04} found strong modulations of SdHO 
in an ultra-clean 2D sample subjected to microwaves of 146\,GHz, 
clearly showing, in addition to the first node 
at $\omega/\omega_c=1$, higher order nodes around $\omega/\omega_c=2$ and 3. 
Mani\cite{Mani-apl} reported strong modulation in the amplitude of SdHOs 
accompanying MIMOs and zero-resistance states excited by a 163.5-GHz radiation 
and large dropoff of the dissipative resistivity below its dark value at
high ($\omega_c/\omega > 1$) field side when subjected to low-frequency radiation. 
Very recently, Dorozhkin {\it et al.}\cite{Dor04} reported both the strong 
suppression of the magnetoresistance caused by radiation below 30\,GHz 
and an interesting modulation of SdHOs in the range $\omega_c/\omega > 1$. 
They found that SdHOs are generally 
strongly damped by the radiation but there is a narrow magnetic field range 
in between allowed ranges of inter- and intra-Landau level transitions, 
where the amplitude of SdHO is insensitive to the microwave 
irradiation. These observations provide a more complicated and appealing picture 
of the microwave-related transport phenomena, which must be accounted for in any 
theoretical model for MIMOs.  

We propose that these SdHO modulations come from the electron heating
induced by the microwave radiation. Under the illumination of microwave 
the electron system, which continuously absorbs energy from the radiation field, 
would certainly be heated. Unfortunately, the electron heating has so far been 
ignored in most of the theoretical treatments. The electron-acoustic phonon 
interaction was previously considered to contribute to Landau-level broadening\cite{Lei04}
or to act as a damping\cite{Inar} for the orbit movement, providing a mechanism 
for the suppression of MIMOs when the lattice temperature increases. Besides the inelastic
electron-phonon scattering also plays another important role to dissipate energy from
the electron system to the lattice. The energy absorption rate is indeed small 
in high-mobility electron systems at low temperature as in the experiments.
This, however, does not imply a negligible electron heating, since the electron 
energy-dissipation rate is also small because of weak electron-phonon scattering 
at temperature $T\leq 1$\, K.   
To deal with SdHO, which is very 
sensitive to the smearing of the electron distribution, one has to carefully calculate 
the electron heating due to microwave irradiation in a uniform model. 

On the other hand, microwave irradiation heats the electrons and thus greatly 
strengthens the thermalizing trend of the system by 
enhancing the electron-electron scattering rate 
at this low temperature regime. This enables us to describe these high-mobility 
2D electron systems with a quasi-equilibrium distribution in a moving reference frame. 

In this paper we pursue a theoretical investigation on MIMOs and SdHOs taking account of 
the electron heating under microwave irradiation. We generalize the balance equation 
approach to radiation-induced magnetotransport in high mobility two-dimensional 
electron systems. By carefully calculating the electron heating 
based on the balance of the energy absorption from the radiation 
field and the energy dissipation to the lattice through electron-phonon 
interactions in a typical GaAs-based heterosystem and taking into account the 
electrodynamic effect, we are able not only to reproduce the interesting 
phenomena of MIMOs in quantitative agreement with experiments in amplitudes, 
phases and radiation dependence of the oscillation, but also to obtain SdHO modulations 
observed in the experiments.

\section{Formulation}
\subsection{Force- and energy-balance equations}

This paper is concerned with the magnetotransport in a microscopically 
homogeneous 2D system, and refers the measured zero resistance to  
the macroscopic consequence of the instability  
due to the occurrence of negative dissipative resistivity. 
     
We consider $N_{\rm e}$ electrons in a unit area of an infinite quasi-2D system
in the $x$-$y$ plane with a confining potential $V(z)$ in the $z$ direction. 
These electrons, in addition to interacting with each other, are scattered by
random impurities and/or disorders and by phonons in the lattice. 
Within the magnetic field range relevant to MIMO phenomenon, the experiments 
exclude the onset of the quantum Hall effect, thus allowing us to assume that the 2D
electrons are in extended states.
  
To include possible elliptically polarized microwave illumination we assume that 
a dc electric field ${\bf E}_0$ and a high-frequency (HF) ac field 
of angular frequency $\omega$,
\begin{equation}
{\bf E}(t)\equiv{\bf E}_s \sin(\omega t)+{\bf E}_c\cos(\omega t),\label{Et}
\end{equation} 
 are applied inside the 2D system in the $x$-$y$ plane, 
together with a magnetic field ${\bf B}=(0,0,B)$ along the $z$ direction.
The spatial homogeneity of the fields and the parabolic band structure allows 
to describe the transport of this system in terms of its center-of-mass (c.m.) motion
and the relative motion, i.e. the motion of electrons in the reference frame 
moving with the c.m.\cite{Ting,Lei85,Lei851}
The center-of-mass momentum and coordinate of
the 2D electron system are defined as 
${\bf P}\equiv\sum_j {\bf p}_{j\|}$ 
and ${\bf R}\equiv N_{\rm e}^{-1}\sum_j {\bf r}_{j\|}$  
with ${\bf p}_{j\|}\equiv(p_{jx},p_{jy})$ and ${\bf r}_{j\|}\equiv (x_j,y_j)$
being the momentum and coordinate of the $j$th electron in the 2D plane, respectively,
and the relative electron momentum and coordinate are defined as
${\bf p}_{j\|}'\equiv{\bf p}_{j\|}-{\bf P}/N_{\rm e}$ and 
${\bf r}_{j\|}'\equiv{\bf r}_{j\|}-{\bf R}$, respectively.
In terms of these variables, the Hamiltonian of the system, $H$, can be written 
as the sum of a center-of-mass part $H_{\rm cm}$
and a relative electron part $H_{\rm er}$ 
(${\bf A}({\bf r})$ is the vector potential of the ${\bf B}$ field),
\begin{eqnarray}
H_{\rm cm}&=&\frac 1{2N_{\rm e}m}({\bf P}-N_{\rm e}e{\bf A}({\bf
R}))^2-N_{\rm e}e({\bf E}_{0}+{\bf E}(t))\cdot {\bf R},\,\,\label{Hcm}\\
H_{\rm er}&=&\sum_{j}\Big[\frac{1}{2m}\left({\bf p}_{j\|}'-e{\bf A}
({\bf r}_{j\|}')\right)^{2}
+\frac{p_{jz}^2}{2m_z}+V(z_j)\Big]\nonumber\\
&&\hspace*{2cm}+\sum_{i<j}V_c({\bf r}_{i\|}'-{\bf r}_{j\|}',z_i,z_j),\label{Her}
\end{eqnarray}
together with electron-impurity and electron-phonon interactions
\begin{eqnarray}
H_{\rm ei}&=&\sum_{j,a,{\bf q}_{\|}} u({\bf q}_{\|},z_a)\,e^{i{\bf q}_{\|}
\cdot({\bf R}+{\bf r}'_{j{\|}}-{\bf r}_{a{\|}})},\\
H_{\rm ep}&=&\sum_{j,{\bf q}_{\|}} M({\bf q}_{\|},q_z)(b_{\bf q}+b_{\bf q}^{\dag})
\,e^{i{\bf q}_{\|}\cdot({\bf R}+{\bf r}'_{j{\|}})}.
\end{eqnarray}
 Here $m$ and $m_z$ are, respectively, the electron effective mass
parallel and perpendicular to the 2D plane, and $V_c$ stands for the electron-electron
Coulomb interaction;  $u({\bf q}_{\|},z_a)$ is the potential of the $a$th impurity
locating at $({\bf r}_{a{\|}},z_a)$; $b_{\bf q}^{\dag}(b_{\bf q})$ are the creation 
(annihilation) operators of the bulk phonon with wavevector ${\bf q}=({\bf q}_{\|},q_z)$ 
and $M({\bf q}_{\|},q_z)$ is the matrix element of the electron-phonon interaction
in the 3D plane-wave representation. 
Note that the uniform electric field (dc and ac) appears only in 
$H_{\rm cm}$, and that $H_{\rm er}$ is just the Hamiltonian of a quasi-2D system 
subjected to a magnetic field without an electric field.
The coupling between the center-of-mass and the relative electrons appears only in 
the exponential factor $\exp(i{\bf q}_{\|}\cdot{\bf R})$ inside the 2D momemtum 
${\bf q}_{\|}$ summation in $H_{\rm ei}$ 
and $H_{\rm ep}$.\cite{Lei851} The balance equation treatment starts with
the Heisenberg operator equation for the rate of change of the center-of-mass velocity 
$\dot{\bf V}=-i[{\bf V},H]+\partial{\bf V}/\partial t$ with ${\bf V}=-i[{\bf R},H]$,
and that for the rate of change of the relative electron energy 
$\dot{H}_{\rm er}=-{\rm i}[H_{\rm er},H]$. Then we proceed with the determination of their
statistical averages.

As proposed in Ref.\,\onlinecite{Lei85}, the c.m. coordinate operator 
${\bf R}$ and velocity operator ${\bf V}$ can be treated classically, i.e. as the 
time-dependent expectation values of c.m. coordinate and velocity,
${\bf R}(t)$ and ${\bf V}(t)$, such that ${\bf R}(t)-{\bf R}(t^{\prime})
=\int_{t^{\prime}}^t{\bf V}(s)ds$.
We are concerned with the steady transport state
under an irradiation of single frequency and focus on the 
photon-induced dc resistivity and the energy absorption of the HF field. 
These quantities are directly related to the time-averaged and/or base-frequency 
oscillating components of the c.m. velocity.
Although higher harmonics of the current 
may affect the dc and lower harmonic terms of the drift velocity 
through entering the damping force and energy exchange rates 
in the resulting equations, in an ordinary semiconductor the power of 
even the third harmonic current is rather weak as compared to the fundamental current. 
For the HF field intensity in the MIMO experiments, 
the effect of higher harmonic current is safely negligible. 
Hence, it suffices to assume that the c.m. 
velocity, i.e. the electron drift velocity, consists of a dc
part ${\bf v}_0$ and a stationary time-dependent part ${\bf v}(t)$ of the form
\begin{equation}
{\bf V}(t)={\bf v}_0-{\bf v}_1 \cos(\omega t)-{\bf v}_2 \sin(\omega t).
\end{equation}
With this, the exponential factor in the operator equations can be expanded 
in terms of Bessel functions ${J}_n(x)$,
\begin{eqnarray}
\hspace*{-1.3cm}&&{\rm e}^{{\rm i}{\bf q}_{\|}\cdot \int_{t^{\prime }}^{t}{\bf V}(s)ds}
=\sum_{n=-\infty }^{\infty }{J}_{n}^{2}(\xi ){\rm e}^{{\rm i}({\bf q}_{\|}\cdot 
{\bf v}_0-n\omega) (t-t^{\prime
})}+\nonumber\\
&&\sum_{m\neq 0}{\rm e}^{{\rm i}m(\omega t-\varphi )}\sum_{n=-\infty }^{\infty
}{J}_{n}(\xi ){J}_{n-m}(\xi ){\rm e}^{{\rm i}({\bf q}_{\|}\cdot 
{\bf v}_0-n\omega) (t-t^{\prime })}.\nonumber
\end{eqnarray}
Here the argument in the Bessel functions 
\begin{equation}
\xi\equiv \frac{1}{\omega}\left[({\bf q}_\|\cdot {\bf v}_1)^2+
({\bf q}_\|\cdot {\bf v}_2)^2\right]^{\frac{1}{2}},
\end{equation}
and  
$\tan \varphi=({\bf q}_{\|}\cdot {\bf v}_2)/({\bf q}_{\|}\cdot {\bf v}_1)$.

Under the influence of a modest-strength HF electric field the electron system is 
far from equilibrium. However, the distribution function of relative electrons, 
which experience no electric field directly, may be close to an quasi-equilibrium 
type distribution function.
For the experimental GaAs-based ultra-clean 2D electron systems having carrier 
mobility of the order of $2000$\,m$^{2}$/Vs, the elastic momentum scattering rate is around 
$\tau_m^{-1} \sim 10$\,mK. In these systems, the thermalization time $\tau_{\rm th}$ 
(i.e. the time for system to return to its internal equilibrating state 
when it is deviated from), estimated conservatively using electron-electron (e-e) 
interaction related inelastic scattering time $\tau_{ee}$ calculated
with an equilibrium distribution function at temperature $T=1$\,K, is also around 
$\tau_{\rm th}^{-1}\sim\tau_{ee}^{-1} \sim 10$\,mK. The illumination of microwave 
certainly heats the electrons. Even an electron heating comparable to a couple of 
degrees temperature rise would greatly enhance $\tau_{ee}^{-1}$, such that the 
thermalization time $\tau_{\rm th}$ would become much shorter than the momentum 
relaxation time $\tau_m$ under microwave irradiation.\cite{note-life} 
The relative electron systems subject to a modest radiation would rapidly 
thermalize and can thus be described reasonably by a Fermi-type distribution 
function at an average electron temperature $T_{\rm e}$ in the reference frame moving 
with the center-of-mass. This allows us to carry out the statistical average 
of the operator equations for the rates of changes of the c.m. velocity ${\bf V}$ 
and relative electron energy $H_{\rm er}$ to the leading order in $H_{\rm ei}$ 
and $H_{\rm ep}$ with succinct forms.

For the determination of unknown parameter ${\bf v}_0$, ${\bf v}_1$, ${\bf v}_2$, 
and ${T_{\rm e}}$, it suffices to know the damping force 
up to the base frequency oscillating term 
${\bf F}(t)= {\bf F}_0+{\bf F}_s\sin(\omega t)+{\bf F}_c\cos(\omega t)$, 
and the energy-related 
quantities up to the time-average terms. We finally obtain the  
force and energy balance equations,

\begin{eqnarray}
&&N_{\rm e}e{\bf E}_{0}+N_{\rm e} e ({\bf v}_0 \times {\bf B})+
{\bf F}_0=0,\label{eqv0}\\
&&{\bf v}_{1}=\frac{e{\bf E}_s}{m\omega}+\frac{{\bf F}_s}{N_{\rm e}m\omega }
-\frac{e}{m\omega }({\bf v}_{2}\times
{\bf B}),\label{eqv1}\\
&-&{\bf v}_{2}=\frac{e{\bf E}_c}{m\omega}+\frac{{\bf F}_c}{N_{\rm e}m\omega }
-\frac{e}{m\omega }({\bf v}_{1}
\times {\bf B}),\label{eqv2}\\
&&N_{\rm e}e{\bf E}_0\cdot {\bf v}_0+S_{\rm p}- W=0.\label{eqsw}
\end{eqnarray}
Here
\begin{eqnarray}
{\bf F}_{0}=\sum_{{\bf q}_\|}\left| U({\bf q}_\|%
)\right| ^{2}%
\sum_{n=-\infty }^{\infty }{\bf q}_\|{J}_{n}^{2}(\xi )\Pi _{2}({\bf %
q}_\|,\omega_0-n\omega )\,\,\,\,\,&&\nonumber\\
+\sum_{{\bf q}}\left| M({\bf q})\right|
^{2}\sum_{n=-\infty
}^{\infty }{\bf q}_\|{J}_{n}^{2}(\xi )\Lambda _{2}({\bf q},\omega_0+
\Omega _{{\bf q}}-n\omega )&&
 \label{eqf0}
\end{eqnarray}
is the time-averaged damping force, 
\begin{eqnarray}
S_{\rm p}=\sum_{{\bf q}_\|}\left| U({\bf q}_\|%
)\right| ^{2}%
\sum_{n=-\infty }^{\infty }n\omega J_{n}^{2}(\xi )\Pi _{2}({\bf %
q}_\|,\omega_0-n\omega )\,\,\,\,\,&&\nonumber\\
+\sum_{{\bf q}}\left| M({\bf q})\right|
^{2}\sum_{n=-\infty
}^{\infty }n\omega J_{n}^{2}(\xi )\Lambda _{2}({\bf q},\omega_0+
\Omega _{{\bf q}}-n\omega )&&
 \label{eqsp}
\end{eqnarray}
is the time-averaged rate of the electron energy absorption from the HF field, and 
\begin{equation}
W=\sum_{{\bf q}}\left| M({\bf q})\right|
^{2}\sum_{n=-\infty
}^{\infty } \Omega_{\bf q}J_{n}^{2}(\xi )\Lambda _{2}({\bf q},\omega_0+
\Omega _{{\bf q}}-n\omega )
 \label{eqw}
\end{equation}
is the time-averaged rate of the electron energy dissipation to the lattice 
due to electron-phonon scatterings. 
The oscillating frictional force amplitudes 
${\bf F}_s\equiv {\bf F}_{22}-{\bf F}_{11}$ and 
${\bf F}_c\equiv {\bf F}_{21}+{\bf F}_{12}$ are given by ($\mu=1,2$)
%\begin{widetext}
\begin{eqnarray}
{\bf F}_{1\mu}=-\sum _{{\bf q}_\|}{\bf q}_\|\eta_{\mu}| U({\bf q}_\|)| ^{2}
\sum_{n=-\infty }^{\infty }\left[ {J}_{n}^{2}(\xi )\right] ^{\prime }\Pi _{1}
({\bf q}_\|,\omega_0-n\omega )
&&\nonumber\\
- 
\sum_{\bf q}{\bf q}_\|\eta_{\mu}| M({\bf q})|
^{2}\sum_{n=-\infty
}^{\infty }\left[ {J}_{n}^{2}(\xi )\right] ^{\prime }\Lambda _{1}({\bf q}, 
\omega_0+\Omega _{{\bf q}}-n\omega ),\,\,\,\,&&\label{eqf1u}\\ 
{\bf F}_{2\mu}=\sum_{{\bf q}_\|}{\bf q}_\|\frac{\eta_{\mu}}
{\xi}| U({\bf q}_\|)| ^{2}
\sum_{n=-\infty }^{\infty }2n{J}_{n}^{2}(\xi )\Pi _{2}({\bf q}_\|,\omega_0-n\omega )
\,\,\,\,\,\,&&\nonumber\\
+ 
\sum_{{\bf q}}{\bf q}_\|\frac{\eta_{\mu}}{\xi}| M({\bf q})|^{2}\sum_{n=-\infty
}^{\infty }2n{J}_{n}^{2}(\xi )\Lambda _{2}({\bf q},\omega_0+\Omega _{\bf q}-n\omega ).
\,\,\,\,\,\,&&\label{eqf2u}
\end{eqnarray}
%\end{widetext}
In these expressions,
$\eta_{\mu}\equiv {\bf q}_\|\cdot {\bf v}_{\mu}/\omega \xi$;
$\omega_0\equiv {\bf q}_\|\cdot {\bf v}_0$;
$U({\bf q}_\|)$ and $M({\bf q})$ are effective impurity and phonon
scattering potentials (including effects of the spatial distribution of
impurities and the form factor of quasi-2D electrons).\cite{Lei851}  
$\Pi_2({\bf q}_\|,\Omega)$ and
$
\Lambda_2({\bf q},\Omega)=2\Pi_2({\bf q}_\|,\Omega)
[n(\Omega_{\bf q}/T)-n(\Omega/T_{\rm e})]
$\,(with $n(x)\equiv 1/({\rm e}^x-1)$)
are the imaginary parts of the electron density correlation function 
and electron-phonon correlation function in the presence of the magnetic field.
$\Pi_1({\bf q}_\|,\Omega)$ and  $\Lambda_1({\bf q},\Omega)$
are the real parts of these two correlation functions.

Effects of a microwave radiation on electron transport first come from 
the HF field induced c.m. motion (electron drift motion) and the related
change of the electron distribution. In addition to this,
the HF field also enters via the argument $\xi$ of the Bessel functions 
in ${\bf F}_0$, ${\bf F}_{\mu\nu}$, $W$ and $S_{\rm p}$. 
Compared with that without a HF field, 
we see that in an electron gas having impurity and/or phonon scatterings 
(otherwise homogeneous),
a HF field of frequency $\omega$ opens additional channels for electron 
transition: an electron in a state can absorb or emit one or several photons of 
frequency $\omega$
and scattered to a different state with the help of impurities and/or phonons.
The sum over $|n|\geq 1$ represents contributions of real single and multiple
photon participating processes. The role of these processes is two folds. 
On the one hand, they contribute additional damping force to the moving electrons, 
giving rise directly to photoresistance, and at the same time, 
transfer energy from the HF field to the electron system, resulting in electron heating, 
i.e. another change (smearing) in the electron distribution.\cite{Lei98} 
Furthermore, the radiation field, showing up 
in the term with $J_0(\xi)$ in ${\bf F}_0$, ${\bf F}_{\mu\nu}$ and $W$, gives 
rise to another effective change of damping forces and energy-loss rate,
without emission or absorption of real photons. This virtual photon process
also contributes to photoresistance.\cite{Lei-apl} 
All these effects are carried by parameters ${\bf v}_0$, ${\bf v}_1$, ${\bf v}_2$ 
and $T_{\rm e}$. 
Eqs.\,(\ref{eqv0})-(\ref{eqsw}) form a closed set of equations for the 
determination of these parameters when ${\bf E}_0$, ${\bf E}_c$ and ${\bf E}_s$ 
are given in a 2D system subjected to a magnetic field $B$ at temperature $T$. 

\subsection{Longitudinal and transverse resistivities}
The nonlinear resistivity in the presence of a high-frequency
field is easily obtained from Eq.\,(\ref{eqv0}).
Taking ${\bf v}_0$ to be in the $x$ direction,
${\bf v}_0=(v_{0x},0,0)$, we immediately get the
transverse and longitudinal resistivities, 
\begin{eqnarray}
R_{xy}&\equiv &\frac{E_{0y}}{N_{\rm e}ev_{0x}}=\frac{B}{N_{\rm e}e},\label{rxy}\\ 
R_{xx}&\equiv &\frac{E_{0x}}{N_{\rm e}ev_{0x}}
=-\frac{F_0}{N_{\rm e}^2 e^2 v_{0x}}.\label{rxx}
\end{eqnarray}
The linear magnetoresistivity is 
the weak dc current limit ($v_{0x}\rightarrow 0$): 
\begin{eqnarray}
R_{xx}&=&-\sum_{{\bf q}_\|}q_x^2\frac{|
U({\bf q}_\|)| ^2}{N_{\rm e}^2 e^2}\sum_{n=-\infty }^\infty {J}_n^2(\xi)\left. 
\frac {\partial \Pi_2}{\partial\, \Omega }\right|_{\Omega =n\omega }\nonumber\\
&&- \sum_{ {\bf q}} q_x^2\frac{\left| M ( {\bf
q})\right| ^2}{N_{\rm e}^2 e^2}\sum_{n=-\infty }^\infty {J}_n^2(\xi)\left. 
\frac {\partial \Lambda_2}{\partial\, \Omega }\right|_{\Omega =\Omega_{{\bf q}}+n\omega}.
\,\,\,\,\,\,\,\label{lrxx}
\end{eqnarray}

Note that although according to Eqs.\,(\ref{eqf0}), (\ref{rxx}) and (\ref{lrxx}), the 
longitudinal magnetoresistivity $R_{xx}$ can be formally written 
as the sum of contributions from various individual scattering mechanisms,
all the scattering mechanisms have to be taken into account simultaneously 
in solving the momentum- and energy-balance equations (\ref{eqv1}), (\ref{eqv2})
and (\ref{eqsw}) for ${\bf v}_1$, ${\bf v}_2$ and $T_{\rm e}$,
which enter the Bessel functions and other parts in the expression of $R_{xx}$.

\subsection{Landau-level broadening}

In the present model the effects of interparticle Coulomb screening are 
included in 
the electron complex density correlation function $\Pi({\bf q}_{\|},\Omega)
=\Pi_1({\bf q}_{\|},\Omega)+i\Pi_2({\bf q}_{\|},\Omega)$,
which, in the random phase approximation, can be expressed as
\begin{equation}
\Pi({\bf q}_{\|},\Omega)=\frac{\Pi_0({\bf q}_{\|},\Omega)}{\epsilon({\bf q}_\|,\Omega)},
\end{equation}
where
\begin{equation}
\epsilon({\bf q}_\|,\Omega)\equiv 1-V(q_{\|})\Pi_0({\bf q}_{\|},\Omega)
\end{equation}
is the complex dynamical dielectric function, 
\begin{equation}
V(q_{\|})=\frac{e^2}{2\epsilon_0\kappa q_{\|}}H(q_{\|})
\end{equation}
is the effective Coulomb potential with $\kappa$ the dielectric constant of the material 
and $H(q_{\|})$ is a 2D wavefunction-related overlapping integration,\cite{Lei851} 
$\Pi_0({\bf q}_{\|},\Omega)=\Pi_{01}({\bf q}_{\|},\Omega)+i\Pi_{02}({\bf q}_{\|},\Omega)$ 
is the complex density correlation function of the independent electron 
system in the presence
of the magnetic field. With this dynamically screened density correlation function 
the collective plasma modes of the 2DES are incorporated. Disregard these 
collective modes one can just use a static screening 
$\epsilon({\bf q}_{\|},0)$ instead. 

The $\Pi_{02}({\bf q}_{\|}, \Omega)$ function of a 2D
system in a magnetic field can be written in terms of Landau representation:\cite{Ting}
\begin{eqnarray}
&&\hspace{-0.7cm}\Pi _{02}({\bf q}_{\|},\Omega ) =  \frac 1{2\pi
l_{\rm B}^2}\sum_{n,n'}C_{n,n'}(l_{\rm B}^2q_{\|}^2/2)  
\Pi _2(n,n',\Omega),
\label{pi_2}\\
&&\hspace{-0.7cm}\Pi _2(n,n',\Omega)=-\frac2\pi \int d\varepsilon
\left [ f(\varepsilon )- f(\varepsilon +\Omega)\right ]\nonumber\\
&&\,\hspace{2cm}\times\,\,{\rm Im}G_n(\varepsilon +\Omega){\rm Im}G_{n'}(\varepsilon ),
\end{eqnarray}
where $l_{\rm B}=\sqrt{1/|eB|}$ is the magnetic length,
\begin{equation}
C_{n,n+l}(Y)\equiv n![(n+l)!]^{-1}Y^le^{-Y}[L_n^l(Y)]^2\label{cnn}
\end{equation}
with $L_n^l(Y)$ the associate Laguerre polynomial, $f(\varepsilon
)=\{\exp [(\varepsilon -\mu)/T_{\rm e}]+1\}^{-1}$ the Fermi distribution
function, and ${\rm Im}G_n(\varepsilon )$ is the imaginary part of the 
electron Green's function, or the density of states (DOS), of the Landau level $n$.
The real part function $\Pi_{01}({\bf q}_{\|},\Omega)$ and corresponding
$\Lambda_{01}({\bf q}_{\|},\Omega)$ function can be derived from their  
imaginary parts via the Kramers-Kronig relation.

In principle, to obtain the Green's function ${\rm Im}G_n(\varepsilon )$,
a self-consistent calculation has to be carried out from the Dyson equation 
for the self-energy with all the impurity, phonon and e-e scatterings 
included. The resultant $G_n$ is generally a complicated function of
the magnetic field, temperature, and Landau-level index $n$, also
dependent on the different kinds of scatterings. Such a calculation is beyond
the scope of the present study.
In this paper we model the DOS function with a Gaussian-type form ($\varepsilon_n$
is the energy of the $n$-th Landau level):\cite{Ando82,Raikh93}
\begin{equation}
{\rm Im}G_n(\varepsilon)=-\frac{\sqrt{2\pi}}{\Gamma}
\exp\Big[-\frac{2(\varepsilon-\varepsilon_n)^2}{\Gamma^2}\Big]\label{gauss}
\end{equation}
with a broadening width given by  
\begin{equation}
\Gamma=\Big(\frac{8e\omega_c\alpha}{\pi m \mu_0}\Big)^{1/2},\label{gamma}
\end{equation}
where $\mu_0$ is the linear mobility  
in the absence of the magnetic field and $\alpha$ is a semiempirical 
parameter to take into account the difference of 
the transport scattering time $\tau_m$ determining the mobility $\mu_0$,    
from the single particle lifetime $\tau_s$ related to Landau level broadening. The latter 
depends on elastic scatterings of different types and their relative strengths, as well as 
contributions of electron-phonon and electron-electron scatterings.
$\alpha$ will be served as the only adjustable parameter in the present investigation.
 Unlike the semielliptic function, which can model only separated
Landau-level case, a Gaussian-type broadening function can reasonably cover both the 
separated-level and overlapping-level regimes.

\subsection{Effect of radiative decay}

The HF electric field ${\bf E}(t)$ appearing in Eqs.\,(8) and (9) 
is the total (external and induced) field really acting on the 2D electrons. 
Experiments are always performed under the condition of giving external radiation. 
In this paper we assume that the electromagnetic wave is incident perpendicularly 
(along $z$-axis) upon 2DEG from the vacuum with the incident electric field of
\begin{equation} 
{\bf E}_{\rm i}(t)={\bf E}_{{\rm i}s}\sin(\omega t)+ {\bf E}_{{\rm i}c}\cos(\omega t)
\end{equation}
at plane $z=0$. The relation between ${\bf E}(t)$ and ${\bf E}_{\rm i}(t)$
is easily obtained by solving the Maxwell equations connecting both sides of the 2DEG
which is carrying a sheet current density $N_{\rm e}e{\bf v}(t)$.
If the 2DEG locates under the surface plane at $z=0$ of a thick 
(treated as semi-infinite) semiconductor substrate having a refraction index $n_s$, 
we have\cite{Chiu,Liu}
\begin{equation}
{\bf E}(t)=\frac{N_{\rm e}e\,{\bf v}(t)}{(1+n_s)\epsilon_0 c}+
\frac{2}{1+n_s}{\bf E}_{\rm i}(t).\label{thick} 
\end{equation}
If the 2DEG is contained in a thin sample suspended in vacuum at the plane $z=0$,
then
\begin{equation}
{\bf E}(t)=\frac{N_{\rm e}e{\bf v}(t)}{2\epsilon_0 c}+
{\bf E}_{\rm i}(t).\label{thin}
\end{equation}
In the numerical calculation of this paper we consider the latter case and use Eq.\,(\ref{thin})
for the total selfconsistent field ${\bf E}(t)$ in Eqs.\,(\ref{eqv1}) and (\ref{eqv2}).
This electrodynamic effect,\cite{Chiu,Liu} recently refered as radiative decay,\cite{Mikh}
gives rise to an additional damping in the 2DEG response to a given incident
HF field. The induced damping turns out to be much stronger 
than the intrinsic damping due to scattering-related forces ${\bf F}_s$ and ${\bf F}_c$ 
for the experimental high-mobility systems at low temperatures. 
For almost all the cases pertinent to MIMO experiments we can neglect 
the forces ${\bf F}_s$ and ${\bf F}_c$ completely in solving 
${\bf v}_1\equiv(v_{1x},v_{1y})$ and ${\bf v}_2\equiv(v_{2x},v_{2y})$ from  
Eqs.(\ref{eqv1}) and (\ref{eqv2}) for given incident fields 
${\bf E}_{{\rm i}s}$ and ${\bf E}_{{\rm i}c}$, and obtain explicitly
\begin{equation}
\begin{array}{ccc}
v_{1x}&=&(a\chi_{sx}+b\chi_{sy})/\Delta \\
v_{1y}&=&(a\chi_{sy}-b\chi_{sx})/\Delta \\
v_{2x}&=&(-a\chi_{cx}-b\chi_{cy})/\Delta\\
v_{2y}&=&(-a\chi_{cy}+b\chi_{cx})/\Delta
\end{array}
\end{equation}
with $\Delta=(1-\delta_{\omega}^2+\gamma_{\omega}^{2})^2+
(2\gamma_{\omega}\delta_{\omega})^2$,   
and
\begin{eqnarray}
\begin{array}{ccc}
\chi_{sx}&=&\nu_{sx}-\delta_{\omega}\nu_{cy}+\gamma_{\omega}\nu_{cx}\\
\chi_{sy}&=&\nu_{sy}+\delta_{\omega}\nu_{cx}+\gamma_{\omega}\nu_{cy}\\
\chi_{cx}&=&\nu_{cx}+\delta_{\omega}\nu_{sy}-\gamma_{\omega}\nu_{sx}\\
\chi_{cy}&=&\nu_{cy}-\delta_{\omega}\nu_{sx}+\gamma_{\omega}\nu_{sy}
\end{array}
\end{eqnarray}
Here 
\begin{equation}
\nu_{\eta}\equiv-\frac{eE_{{\rm i}\eta}}{m\omega}\,\,\,\,\, (\eta=sx,sy,cx,cy),
\end{equation}
$\delta_{\omega}\equiv\omega_c/\omega$ and $\gamma_{\omega}\equiv\gamma/\omega$ with
\begin{equation}
\gamma=\frac{N_{\rm e}e^2}{2m\epsilon_0 c }.
\end{equation}
With these ${\bf v}_1$ and ${\bf v}_2$, the argument $\xi$ entering the Bessel 
functions is obtained. All the transport quantities, such as $S_{\rm p}$, $W$ 
and $R_{xx}$, can be calculated directly with the electron temperature 
$T_{\rm e}$ determined from the energy balance equation (\ref{eqsw}). 

\section{Numerical results for GaAs-based systems}

As in the experiments, we focus our attention on  
high mobility 2DEGs formed by GaAs/AlGaAs heterojunctions. 
For these systems at temperature $T\leq 1$\,K,
the dominant contributions to the energy absorption $S_{\rm p}$ and photoresistivity 
$R_{xx}-R_{xx}(0)$ come from the impurity-assisted photon-absorption and emission 
process. 
At different magnetic field strength, this process is associated with
electron transitions between either inter-Landau level states or intra-Landau-level
states. According to (\ref{gauss}), the width of each Landau level is about $2\Gamma$. 
The condition 
for inter-Landau level transition with impurity-assisted single-photon 
process\cite{note1}  is $\omega>\omega_c-2\Gamma$, or
$\omega_c/\omega<a_{\rm inter}=(\beta+\sqrt{\beta^2+4})^2/4$;
and that for impurity-assisted intra-Landau level transition is 
$\omega<2\Gamma$, or $\omega_c/\omega>a_{\rm intra}=\beta^{-2}$, 
here $\beta=(32e\alpha/\pi m \mu_0\omega)^{\frac{1}{2}}$. However, 
since the DOS of each Landau level is assumed to be Gaussian rather than a clear 
cutoff function and the multi-photon processes also play roles, the transition 
boundaries between different regimes may be somewhat smeared.

 As indicated by experiments,\cite{Umansky} although long range scattering due to 
remote donors always exists in the 2D heterostructures, in ultra-clean GaAs-based 
2D samples having mobility of order of $10^{3}$\,m$^{2}$/Vs, the remote donor 
scattering is responsible for merely $\sim 10\%$ or less of the total momentum 
scattering rate. The dominant contribution to the momentum scattering rate comes 
from short-range scatterers such as residual impurities or defects in the background.
Furthermore, even with the same momentum scattering rate the remote impurity 
scattering is much less efficient in contributing to microwave-induced magnetoresistance
oscillations than short-ranged background impurities or defects.\cite{Lei0409219}
Therefore, in the numerical calculations in this paper we assume 
that the elastic scatterings are due to short-range impurities randomly distributed
throughout the GaAs region.  The impurity densities are determined by the 
requirement that electron total linear mobility at zero magnetic field equals 
the giving value at lattice temperature $T$.
Possibly, long-range remote donnor scattering may give rise to important 
contribution to the Landau-level broadening. This effect, together with the role of
electron-phonon and electron-electron scatterings, is included in the 
semiempirical parameter $\alpha$ in the expression (\ref{gamma}).
 
In order to obtain the energy dissipation rate from the electron system 
to the lattice, $W$, we take into account
scatterings from bulk longitudinal acoustic (LA) and transverse acoustic (TA) 
phonons (via the deformation 
potential and piezoelectric couplings), as well as from longitudinal optical (LO) 
phonons (via the Fr\"{o}hlich coupling) in the GaAs-based system. 
The relevant matrix elements are well known.\cite{Lei851}
The material and coupling parameters for the system are  
taken to be widely accepted values in bulk GaAs: 
electron effective mass $m=0.068\,m_{\rm e}$ ($m_{\rm e}$ is the
free electron mass), transverse sound speed $v_{\rm st}=2.48\times 10^3$\,m/s,
longitudinal sound speed $v_{\rm sl}=5.29\times 10^3$\,m/s, acoustic 
deformation potential $\Xi=8.5$\,eV, piezoelectric constant $e_{14}=
1.41\times 10^9$\,V/m, dielectric constant $\kappa=12.9$, 
material mass density $d=5.31$\,g/cm$^3$.

\subsection{100 GHz}
 
\begin{figure}
\includegraphics [width=0.475\textwidth,clip=on] {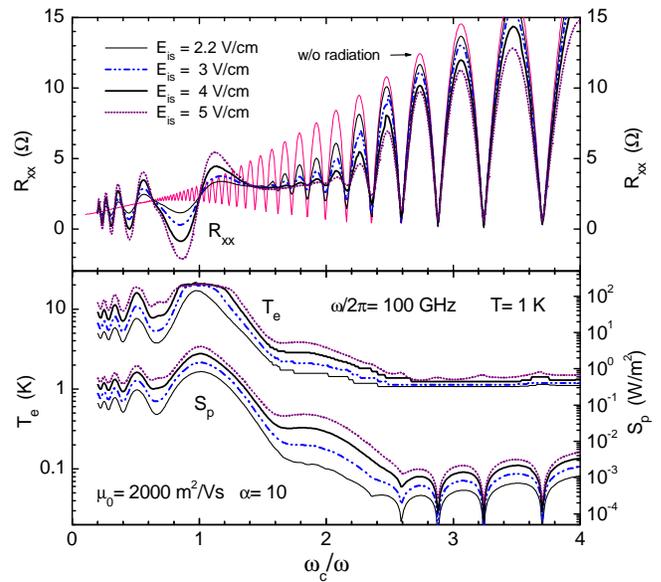}
\vspace*{-0.2cm}
\caption{The magnetoresistivity $R_{xx}$, electron 
temperature $T_{\rm e}$ and energy absorption rate $S_{\rm p}$ of a GaAs-based 2DEG 
with $\mu_0=2000$\,m$^2$/Vs and $\alpha=10$, subjected to 100\,GHz linearly $x$-polarized 
incident HF fields $E_{{\rm i}s}\sin(\omega t)$ of four different strengths.  
The lattice temperature is $T=1 \,K$.}
\label{fig1}
\end{figure}
\begin{figure}
\includegraphics [width=0.42\textwidth,clip=on] {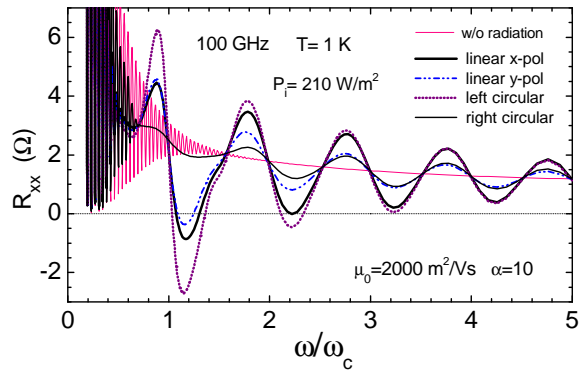}
\vspace*{-0.2cm}
\caption{The magnetoresistivity $R_{xx}$ versus 
$\omega/\omega_c$ for the same system as described in Fig.\,1, subject to 100\,GHz 
HF radiations of fixed incident power of $P_{\rm i}=210$\,W/m$^2$ 
but four different polarizations.}
\label{fig2}
\end{figure}
 
Figure 1 shows the calculated energy absorption rate $S_{\rm p}$, the electron 
temperature $T_{\rm e}$ and the longitudinal magnetoresistivity $R_{xx}$ 
as functions of $\omega_c/\omega$ for a 2D system having an electron density of
$N_{\rm e}=3.0\times 10^{15}$\,m$^{-2}$, a
linear mobility of $\mu_0=2000$\,m$^2$/Vs and a broadening parameter of
$\alpha=10$, subject to linearly $x$-direction polarized incident microwave 
radiations of frequency $\omega/2\pi=100$\,GHz having four different amplitudes 
$E_{{\rm i}s}=2.2, 3, 4$ and 5\,V/cm 
at a lattice temperature of $T=1$\,K.
The energy absorption rate $S_{\rm p}$ exhibits a broad main peak at cyclotron 
resonance $\omega_c/\omega=1$ and secondary peaks at harmonics 
$\omega_c/\omega=1/2,1/3,1/4$. The electron heating has similar feature:
$T_{\rm e}$ exhibits peaks around $\omega_c/\omega=1,1/2,1/3,1/4$. 
For this GaAs system $\beta=0.65$, $a_{\rm inter}=1.6$
and $a_{\rm intra}=4.7$. We can see that, at lower magnetic fields, especially
$\omega_c/\omega<1.4$,  the system absorbs enough energy from the radiation 
field via inter-Landau level transitions and $T_{\rm e}$ is significantly 
higher than $T$, 
with the maximum as high as 21\,K around $\omega_c/\omega=1$. 
With increasing strength of the magnetic field the inter-Landau level transition 
weakens (impurity-assisted single-photon process is mainly allowed when 
$\omega_c/\omega<a_{\rm inter}=1.6$) and the absorbed energy decreases rapidly. 
Within the range $2<\omega_c/\omega<4$ before intra-Landau level transitions 
can take place, $S_{\rm p}$ is two orders of magnitude smaller than that 
in the low magnetic field range. 
Correspondingly the electron temperature $T_{\rm e}$
is only slightly higher than the lattice temperature $T$. 
The magnetoresistivity $R_{xx}$ showing in the upper part of Fig.\,1, exhibits
interesting features.  MIMOs (with fixed points rather than extrema at 
$\omega_c/\omega=1,1/2,1/3,1/4$) clearly appear at lower magnetic fields, which are  
insensitive to the electron heating even at $T_{\rm e}$ of order of 20\,K. 
SdHOs appearing in the higher magnetic field side, however, are damped due to the 
rise of the electron temperature $T_{\rm e}>1$\,K as compared with that without
radiation. With an increase in the microwave amplitude from $E_{{\rm i}s}=2.2$\,V/cm to 
$5$\,V/cm, MIMOs become much stronger and SdHOs are further damped. 
But the radiation-induced SdHO damping is always relatively smaller within 
$2.4<\omega_c/\omega<4$ between allowed ranges of inter- and intra-Landau 
level transitions.

It is worth noting that the predicted MIMOs here exhibit much improved 
agreement with experiments over previous theoretical models. 
The maxima of $R_{xx}$ oscillation locate at $\omega/\omega_c=j-\delta_{-}$ and minima
at $\omega/\omega_c=j+\delta_{+}$, with $\delta_{\pm}\sim 0.23-0.25$ for $j=2,3,4...$
and  $\delta_{\pm}\sim 0.16-0.18$ for $j=1$ (see Fig.\,2). These phase details, as well as
the absolute (rather than reduced) magnitudes of the oscillation amplitudes 
and the required incident microwave strengths to induce oscillations
are in good quantitative agreement with experiments.\cite{Mani,Zud03,Yang,
Dor,Mani04}

The MIMOs depend on the polarization of the incident 
microwave field in respect to the dc field ${\bf E}_0$. 
Physically this is clear in the present model since it is through the c.m. motion
that a HF field affects the photoresistivity of the 2D electron system.
Under the influence of a magnetic field perpendicular to the plane, the c.m.
performs a cyclic motion of frequency $\omega_c$ in the 2D plane. A perpendicularly 
incident circularly-polarized  microwave would accelerate or decelerate this cyclic 
motion depending on the HF electric field circling with or against it. 
Thus, at fix incident power, a left-polarized microwave would yield much stronger effect 
on the $R_{xx}$ oscillation than a right-polarized one and this effect is apparently 
strongest in the vicinity of cylcotron resonance $\omega_c/\omega=1$. 
The difference between the $x$-direction linearly polarized wave and 
the $y$-direction linearly polarized wave, however, comes mainly from the 
the different angle of radiation-induced c.m. motion with respect to the dc current, 
and thus not so sensitive to that of the $\omega_c/\omega$ range. 
In Fig.\,2 we plot the calculated $R_{xx}$ versus 
$\omega/\omega_c$ for the same system as described in Fig.\,1, subject to a 100\,GHz 
microwave radiation having a fixed incident power of $P_{\rm i}=210$\,W/m$^2$ 
(equivalent to an incident amplitude $E_{{\rm i}s}=4$\,V/cm of linear polarization)
but four different polarizations: linear $x$-polarizaton, linear $y$-polarization,
left circular polarization and right circular polarization. 
Their difference is clearly seen.

\subsection{50 GHz and lower frequency}

\begin{figure}
\includegraphics [width=0.475\textwidth,clip=on] {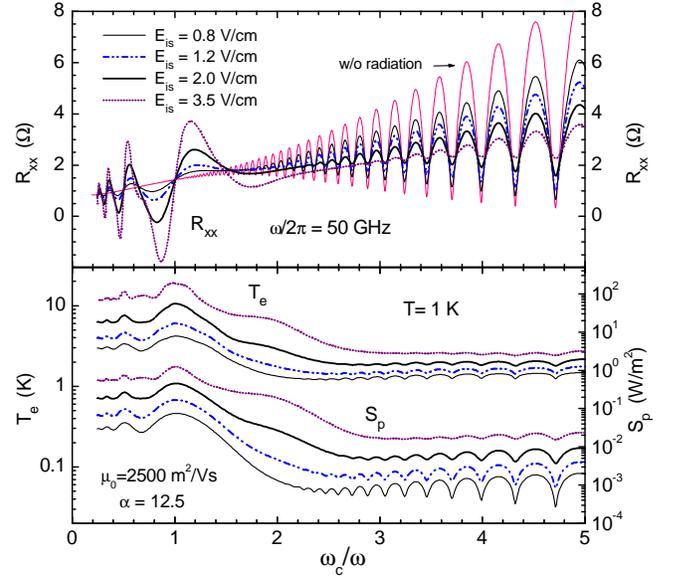}
\vspace*{-0.2cm}
\caption{The magnetoresistivity $R_{xx}$, electron 
temperature $T_{\rm e}$ and energy absorption rate $S_{\rm p}$ of a GaAs-based 2DEG 
with $\mu_0=2500$\,m$^2$/Vs and $\alpha=12.5$, subjected to 50\,GHz linearly $x$-polarized 
incident HF fields $E_{{\rm i}s}\sin(\omega t)$ of four different strengths. 
The lattice temperature is $T=1 \,K$.}
\label{fig3}
\end{figure}

\begin{figure}
\includegraphics [width=0.42\textwidth,clip=on] {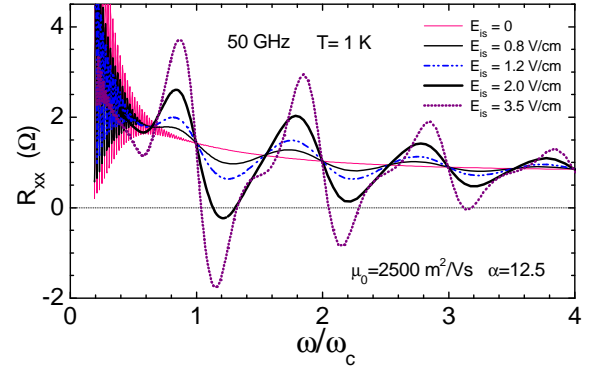}
\vspace*{-0.2cm}
\caption{The magnetoresistivity $R_{xx}$ versus 
$\omega/\omega_c$ for the same system as described in Fig.\,3, subject to 50GHz 
linearly $x$-polarized incident HF fields of four different strengths.}
\label{fig4}
\end{figure}

Figure 3 shows the energy absorption rate $S_{\rm p}$, the electron 
temperature $T_{\rm e}$ and the longitudinal magnetoresistivity $R_{xx}$ 
as functions of $\omega_c/\omega$ for a 2D system having an electron density 
of $N_{\rm e}=3.0\times 10^{15}$\,m$^{-2}$, 
a linear mobility of $\mu_0=2500$\,m$^2$/Vs and a broadening parameter of
$\alpha=12.5$, subject to linearly $x$-direction polarized incident microwave radiations 
of frequency $\omega/2\pi=50$\,GHz having four different amplitudes 
$E_{{\rm i}s}=0.8,1.2,2.0$ and 3.5\,V/cm at a lattice temperature of $T=1$\,K. 
For this GaAs system at 50\,GHz $a_{\rm inter}=1.9$ and $a_{\rm intra}=2.4$. 
The intra-Landau level single-photon transitions are allowed when $\omega_c/\omega > 2.4$, 
yielding, at the high $\omega_c/\omega$ side, an absorption rate $S_{\rm p}$ somewhat larger, 
an electron temperature $T_{\rm e}$ somewhat higher, and a SdHO damping stronger 
than those in the 100-GHz case (Fig.\,1). On the other hand, at equivalent HF field strength
the multiphoton processes are more important at lower frequency. This helps to
enhance the absorption $S_{\rm p}$ in the range $1.9 < \omega_c/\omega < 2.4$, 
where the single-photon process is forbidden and to increase the two-photon 
resonance in $S_{\rm p}$ and $T_{\rm e}$ around $\omega/\omega_c=1.5,2.5$ 
and 3.5 (see $S_{\rm p}$ and $T_{\rm e}$ curves corresponding to 
$E_{{\rm i}s}=3.5$\,V/cm in Fig.\,3). The effect of the two-photon process can also be seen clearly
in the $R_{xx}$-vs-$\omega/\omega_c$ curves as shown in Fig.\,4, where  
the $R_{xx}$ curve of $E_{{\rm i}s}=3.5$\,V/cm exhibits obvious shoulders around 
$\omega/\omega_c=1.5,2.5$ and 3.5, and the descends down around $\omega/\omega_c=0.6$.
This kind of two-photon process was clearly seen in the experiments.\cite{Zud03,Zud04}

\begin{figure}
\includegraphics [width=0.42\textwidth,clip=on] {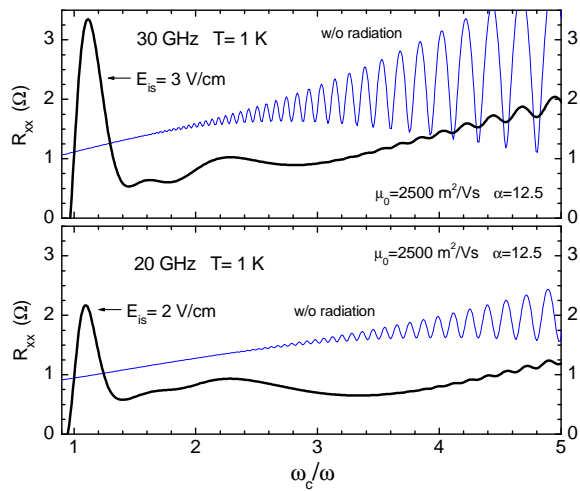}
\vspace*{-0.2cm}
\caption{The magnetoresistivity $R_{xx}$ vs 
$\omega_c/\omega$ for the same system as described in Fig.\,3, subject to  
two incident microwave fields: frequency $\omega/2\pi$=30\,GHz, amplitude
$E_{{\rm i}s}=3$\,V/cm and frequency $\omega/2\pi$=20\,GHz, amplitude
$E_{{\rm i}s}=2$\,V/cm.}
\label{fig5}
\end{figure}

 At even lower frequency, such as 30\,GHz and 20\,GHz, the ranges for 
intra-Landau level and inter-Landau level single-photon transitions 
overlap. The enhanced effect of the virtual photon process,  
together with enhanced multiphoton-assisted electron transition, pushes
the resistivity $R_{xx}$ remarkably down below the average of its oscillatory curve 
without radiation, resulting in a strong suppression of dissipative magnetoresistance 
across a wide magnetic field range as shown in Fig.\,5, in agreement with
experimental observations.\cite{Mani-apl,Dor04}

\subsection{150 and 280 GHz}

The radiation-induced SdHO modulation can be seen clearly in the low magnetic field 
region $\omega/\omega_c>1$ with higher radiation frequency. 
Figure 6 shows the calculated electron 
temperature $T_{\rm e}$ and magnetoresistivity $R_{xx}$ 
as functions of $\omega/\omega_c$ for a 2D system of electron density 
$N_{\rm e}=3.0\times 10^{15}$\,m$^{-2}$, linear mobility  
$\mu_0=2000$\,m$^2$/Vs and $\alpha=3$, subject to a 150-GHz microwave radiation 
of three different amplitudes $E_{{\rm i}s}=0.1, 0.6$ and 2\,V/cm at a lattice 
temperature of $T=0.5$\,K. Low-power
microwave illumination ($E_{{\rm i}s}=0.1$\,V/cm) already yields 
sufficient $T_{\rm e}$ oscillation with maxima at $\omega/\omega_c=1,2,3,4$, 
giving rise to clear SdHO modulations having nodes at $T_{\rm e}$ maxima. At higher 
microwave power ($E_{{\rm i}s}=0.6$\,V/cm) when the MIMO shows up, the $T_{\rm e}$ maxima
gets higher, suppressing the SdHO in the vicinities of $\omega/\omega_c=1,2,3,4$, 
but a strong amplitude modulation of SdHOs is still seen. In the case of $E_{{\rm i}s}=2$\,V/cm, 
$R_{xx}$ shows strong MIMO and the electron temperature further grows so that most of SdHOs 
almost disappear in the range of $\omega/\omega_c>2$. Note that the small $T_{\rm e}$
peaks at $\omega/\omega_c=1.5$ and 2.5 are due to the absorption rate $S_{\rm p}$ maxima 
induced by two-photon processes, which gives rise to additional nodes in the SdHOs. 

\begin{figure}
\includegraphics [width=0.475\textwidth,clip=on] {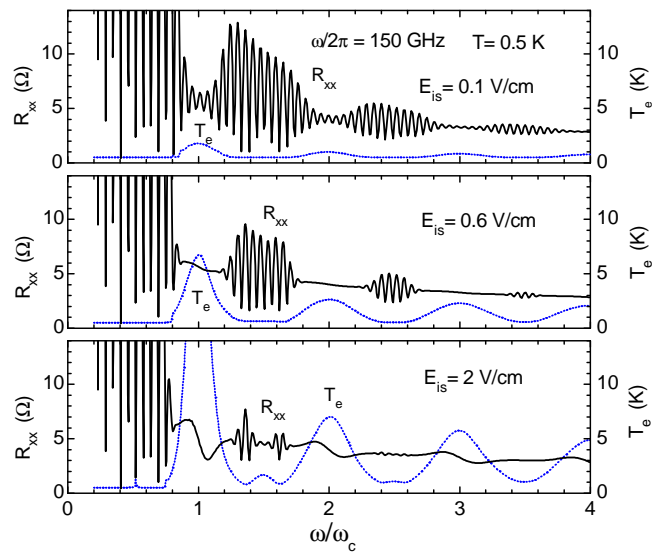}
\vspace*{-0.2cm}
\caption{The magnetoresistivity $R_{xx}$ and electron temperature $T_{\rm e}$ 
of a GaAs-based 2DEG with $N_{\rm e}=3.0\times 10^{15}$\,m$^{-2}$, $\mu_0=2000$\,m$^2$/Vs,
and $\alpha=3$, subjected to 150-GHz linearly $x$-polarized 
incident HF fields $E_{{\rm i}s}\sin(\omega t)$ of three different strengths 
$E_{{\rm i}s}=0.1,0.6$ and 2\,V/cm at $T=0.5$\,K.}
\label{fig6}
\end{figure}
\begin{figure}
\includegraphics [width=0.475\textwidth,clip=on] {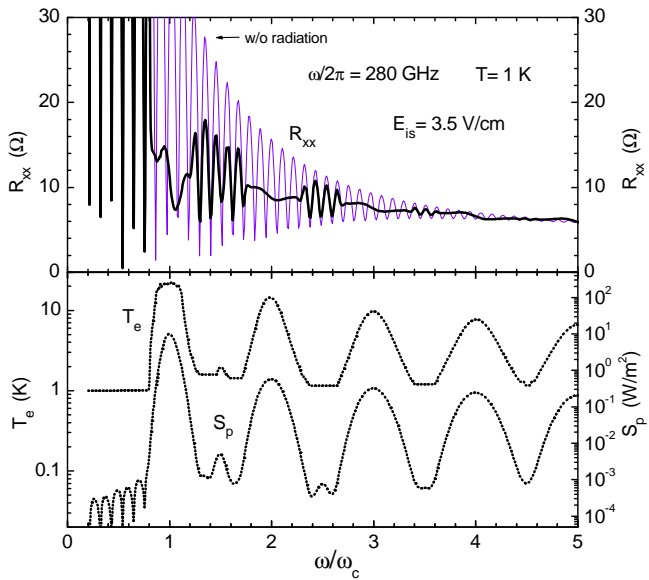}
\vspace*{-0.2cm}
\caption{The magnetoresistivity $R_{xx}$, electron 
temperature $T_{\rm e}$ and energy absorption rate $S_{\rm p}$ of a GaAs-based 2DEG 
with $\mu_0=1000$\,m$^2$/Vs and $\alpha=2$, subjected to a linearly $x$-polarized incident HF field  
of frequency 280\,GHz and amplitude $E_{{\rm i}s}=3.5$\,V/cm at $T=1$\,K.}
\label{fig7}
\end{figure}

Another example of the SdHO modulation appearing simultaneously with MIMO is 
plotted in Fig.\,7, where the energy absorption rate $S_{\rm p}$, the electron 
temperature $T_{\rm e}$, and the magnetoresistivity $R_{xx}$ are shown 
as functions of $\omega/\omega_c$ for a 2D system having an electron density of
$N_{\rm e}=3.0\times 10^{15}$\,m$^{-2}$, a linear mobility of $\mu_0=1000$\,m$^2$/Vs, 
and a broadening parameter of $\alpha=2$, 
subject to linearly $x$-direction polarized incident microwave radiations 
of frequency $\omega/2\pi=280$\,GHz and amplitude $E_{{\rm i}s}=3.5$\,V/cm.
The energy absorption rate $S_{\rm p}$ has broad large peaks 
at $\omega/\omega_c=1,2,3,4,5$ (due to single-photon resonant process) and small peaks 
at $\omega/\omega_c=1.5,2.5$ (due to two-photon resonant process), giving rise to the oscillation
of the electron temperature $T_{\rm e}$. 
One can clearly see 
the peaks of the electron temperature $T_{\rm e}$ and the nodes of SdHO modulation
at $\omega/\omega_c=1,2,3,4$ and 5, together with MIMOs. 
These are in agreement with the experimental observation reported in 
Ref.\onlinecite{Kovalev}.

\subsection{Discussion}         
Note that in GaAs-based systems at a temperature around $T\sim 1$\,K, 
LA phonons generally give larger contribution to the electron energy dissipation $W$ 
than that from TA phonons and LO phonons are usually frozen. However, in the case of 
high radiation power or in the vicinity of $\omega\sim\omega_c$, where the resonantly 
absorbed energy can be relatively large and the electron temperature can rise up
above 20\,K, a weak emission of LO phonons takes place. 
Though at this temperature the number of excited LO phonons is still very small 
and their contribution to momentum relaxation (resitivity) is negligible 
in comparison with acoustic phonons, they can already provide an efficient energy 
dissipation because each excited LO phonon contributes a huge energy transfer of 
$\Omega_{\rm LO}\sim 400$\,K. With a continuing rise of electron temperature the 
LO-phonon contribution increases rapidly. 
This effectively prevents the electron temperature from going much 
higher than 20\,K, such that the $T_{\rm e}$-vs-$\omega_c/\omega$ curve of large 
incident microwave power in Fig.1 exhibits a flat top around $\omega_c/\omega=1$.

In this paper, we did not consider the role of surface or interface phonons 
in the GaAs heterostructure. 
Depending on sample geometry, the surface phonons may be important in dissipating
electron energy thus decreasing the electron temperature.

\vspace*{0.5cm}

\centerline{\bf Acknowledgements}

\vspace*{0.3cm}

We thank Dr. V.I. Ryzhii, Dr. R.G. Mani and Dr. R.R. Du for helpful discussions.
This work was supported by Projects of the National Science Foundation of China, 
the Special Funds for Major State Basic Research Project, 
and the Shanghai Municipal Commission of Science and Technology.


\begin{thebibliography}{99}
\bibitem{note-life} Associated with the elastic collision term in the Boltzmann equation 
for the distribution function is the momenrum relaxation time or transport scattering time
$\tau_m$, which indicates an ability to collect particles along 
a preference direction such as the electric field or motion of c.m., thus inclined
to deviate the system from equilibrating distribution. The  single particle lifetime 
$\tau_s$ related to elastic scattering is the time that a quasiparticle can exist 
in a quantum state before it decays out. It represents an ability of spreading particles 
in all the directions thus towards an isotropic distribution along an equal-energy ring. 
The equilibrating distribution is isotropic in the c.m. reference frame. 
The single particle lifetime $\tau_s$ would not induce any deviation from an 
equilibrating-type distribution.
\bibitem{note1} Mulitiphoton ($|n|>1$) processes of course can take place up to 
higher magnetic fields.
\end{thebibliography}
\end{document}